\documentclass[twoside,11pt]{article}

\usepackage{booktabs}
\usepackage{xcolor}
\usepackage{courier}
\usepackage{float}

\usepackage{jmlr2e}

\hypersetup{
  colorlinks=true,
  linkcolor=blue,
  filecolor=magenta,
  urlcolor=blue,
  citecolor=purple
}

\ShortHeadings{Submitted to Journal of Machine Learning Research (20/10/2020)}{Gros et al}
\firstpageno{1}

\begin{document}

\title{ivado{\color{red}{med}}: A Medical Imaging Deep Learning Toolbox}

\author{\name Charley Gros$^1$ \email charley.gros@gmail.com \vspace{-5pt}\AND 
\name Andreanne Lemay$^1$ \email andreanne.lemay@polymtl.ca \vspace{-5pt}\AND
\name Olivier Vincent$^1$ \email ovincent.poly@gmail.com \vspace{-5pt}\AND
\name Lucas Rouhier$^1$ \email lucasrouhier@gmail.com \vspace{-5pt}\AND
\name Anthime Bucquet$^1$ \email an.bucquet@gmail.com \vspace{-5pt}\AND
\name Joseph Paul Cohen$^2$ \email joseph@josephpcohen.com \vspace{-5pt}\AND
\name Julien Cohen-Adad$^1$ \email jcohen@polymtl.ca \vspace{5pt}\AND
$^1$\addr NeuroPoly Lab, Institute of Biomedical Engineering, Polytechnique Montreal, Montreal, Canada\\
$^2$\addr AIMI, Stanford University and Mila, Quebec AI Institute
}


\maketitle

\begin{abstract}
\texttt{ivadomed} is an open-source Python package for designing, end-to-end training, and evaluating deep learning models applied to medical imaging data. The package includes APIs, command-line tools, documentation, and tutorials. \texttt{ivadomed} also includes pre-trained models such as spinal tumor segmentation and vertebral labeling. Original features of \texttt{ivadomed} include a data loader that can parse image metadata (e.g., acquisition parameters, image contrast, resolution) and subject metadata (e.g., pathology, age, sex) for custom data splitting or extra information during training and evaluation. Any dataset following the Brain Imaging Data Structure (BIDS) convention will be compatible with \texttt{ivadomed} without the need to manually organize the data, which is typically a tedious task. Beyond the traditional deep learning methods, \texttt{ivadomed} features cutting-edge architectures, such as FiLM and HeMis, as well as various uncertainty estimation methods (aleatoric and epistemic), and losses adapted to imbalanced classes and non-binary predictions. Each step is conveniently configurable via a single file. At the same time, the code is highly modular to allow addition/modification of an architecture or pre/post-processing steps. Example applications of \texttt{ivadomed} include MRI object detection, segmentation, and labeling of anatomical and pathological structures. Overall, \texttt{ivadomed} enables easy and quick exploration of the latest advances in deep learning for medical imaging applications. \texttt{ivadomed}'s main project page is available at \url{https://ivadomed.org}. 
\end{abstract}




\section{Introduction}

Deep learning is increasingly used in medical image processing \citep{kim_deep_2019}. It provides automated solutions to repetitive and/or tedious tasks such as the segmentation of pathological structures. However, medical imaging data present many challenges: datasets are often not publicly-available, ground truth labels are scarce due to the limited availability of expert raters, and needs can be very specific and tailored to particular datasets (e.g., segmentation of spinal tumors on sagittal MRI T2-weighted scans). Thus, offering solution for convenient training models (or fine-tuning of pre-existing models) is needed. 

We present \texttt{ivadomed}, a deep learning toolbox dedicated to medical data processing. \texttt{ivadomed} aims to support the integration of deep learning models into the clinical routine, as well as state-of-the-art academic biomedical research. It features intuitive command-line tools for end-to-end training and evaluation of various deep learning models. The package also includes pre-trained models that can be used to accommodate specific datasets with transfer learning.

Another challenge of medical imaging is the heterogeneity of the data across clinical centers, in terms of image features (e.g., contrast, resolution) and population demographics. This makes it challenging to create models that can generalize well across the many existing datasets. Recent cutting-edge methods address this problem, such as FiLM \citep{perez2017film} and HeMis \citep{havaei2016hemis}, however they are usually not implemented within a comprehensive framework that enables end-to-end training and experimentation. In addition to providing these advanced architectures, \texttt{ivadomed} features multiple uncertainty estimation methods (aleatoric and epistemic), losses adapted to imbalanced classes and non-binary predictions. Each step can be conveniently customized via a single configuration file, and at the same time, the code is highly modular to allow addition/modification of architecture or pre-/post-processing steps. 

\texttt{ivadomed} is based on PyTorch framework \citep{paszke2017automatic} with GPU acceleration supported by CUDA. It can easily be installed via \href{https://pypi.org/project/ivadomed/}{PyPI} and the whole package is tested with a \href{https://github.com/ivadomed/ivadomed/actions?query=workflow\%3A\%22Run+tests\%22}{continuous integration} framework. The project website, which includes user and API documentation, is available at \url{https://ivadomed.org}. The name \texttt{ivadomed} is a portmanteau between \textit{IVADO} (\href{https://ivado.ca}{The Institute for data valorization}) and \textit{medical}.

\begin{figure}[h]
\vspace{-5pt}
\includegraphics[width=1\linewidth]{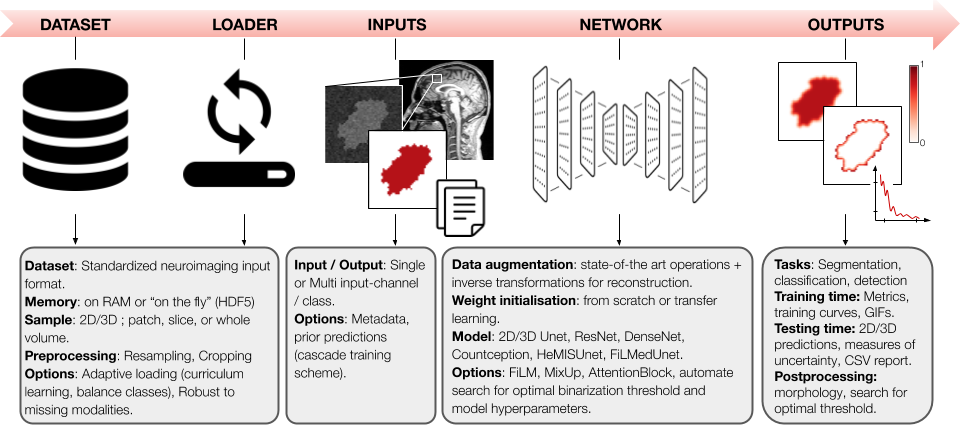}
\caption{Overview of \texttt{ivadomed} main features.}
\label{fig:main}
\vspace{-20pt}
\end{figure}

\section{Software Description}

\paragraph{Loader:}

An important aspect of machine learning is data management. Lots of time is usually spent manually organizing data into a proper structure to make the dataset compatible with a chosen analysis pipeline \citep{Willemink2020-au}. \texttt{ivadomed} features a data loader module that expects datasets to be structured according to a widely-accepted convention: the Brain Imaging Data Structure (BIDS) \citep{bids_2016}. Thus, any dataset following the BIDS convention can immediately be used by \texttt{ivadomed}, e.g., for training a new model, without the need to spend time organizing the data. BIDS convention is designed around neuroimaging MRI data and accepts NIfTI file formats, but the BIDS community is actively expanding its specifications to other modalities (CT, MEG/EEG, microscopy) and file formats (PNG, OME-TIFF), which \texttt{ivadomed} will then be able to accommodate. 

One benefit of the BIDS convention is that each image file is associated with a JSON file containing metadata. \texttt{ivadomed}'s loader can parse image metadata (e.g., acquisition parameters, image contrast, resolution) and subject metadata (e.g., pathology, age, sex) for custom data splitting or extra information during training and evaluation. It is possible to modulate specific layers of a convolutional neural network using metadata information (e.g., image contrast, data center, disease phenotype), to tailor it towards a particular data domain or to enable experiments with architectures such as FiLM \citep{perez2017film} (which is implemented in \texttt{ivadomed}). Metadata could also be useful to mitigate class imbalance via data balancing techniques.

\texttt{ivadomed}'s data loader can accommodate 2D/3D images, multiple input channels as well as multiple prediction classes. Images can be loaded as a volume, slice-wise, or patch-wise. Data can be saved on the RAM or used "on the fly" via HDF5 format. Cropping, resampling, normalization, and histogram clipping and equalization can be applied during the loading as a pre-processing step. \texttt{ivadomed} can deal with missing modalities \citep{havaei2016hemis} by resorting to curriculum learning to train the model. 

\paragraph{Training:}

\texttt{ivadomed} includes all the necessary components for training segmentation models from start to finish. The first step usually consists of applying data augmentation transformations, such as affine, elastic, and ground-truth dilation, which are all included in \texttt{ivadomed}. The next step is model training, which can be done either from scratch or using transfer learning on a pre-trained model by freezing some layers. Available architectures include: 2D U-Net \citep{Ronneberger2015unet}, 3D U-Net \citep{isensee2017brain}, ResNet \citep{he2016resnet}, DenseNet \citep{Huang2017densenet}, Count-ception \citep{Cohen2017countception}, and HeMIS U-Net. These models can easily be enriched via attention blocks \citep{oktay2018attention} or FiLM layers (which modulate U-Net features using metadata). To facilitate the training process, \texttt{ivadomed} offers multiple loss functions such as the Dice coefficient \citep{milletari2016v}, cross-entropy, and L2 norm, including some adapted to medical imaging challenges, such as the adaptive wing loss \citep{wang_adaptive_2019} for soft labels and the focal Dice loss \citep{wong20183d} for class imbalance. To partly address the problem of small datasets, mixup \citep{zhang2017mixup} has been implemented and adapted for segmentation tasks. To mitigate class imbalance, \texttt{ivadomed} supports cascaded architectures. With a single inference, it is possible to narrow down the region of interest via object detection, and then segment a specific structure. In case of interruption during training, all parameters are saved after each epoch so that training can be resumed at any time. 

It can be overwhelming to get started and choose across all the available models, losses, and parameters. \texttt{ivadomed}'s repository includes the script \texttt{ivadomed\_automate\_training} to configure and launch multiple trainings across GPUs. In addition, \texttt{ivadomed} includes tutorials that cover typical training scenarios such as one-class segmentation, cascade of architectures, and uncertainty estimation. 

\paragraph{Evaluation:}

A model can be thoroughly evaluated on the testing set by computing various popular metrics for segmentation, classification, and regression tasks. Slice-wise or patch-wise predictions are reconstructed in the initial space for evaluation and output visualization. \texttt{ivadomed} can produce aleatoric \citep{wang_aleatoric_2019} and/or epistemic \citep{nair_exploring_2018} uncertainty, voxel-wise and/or object-wise \citep{roy_quicknat_2018}, using multiple available measures (e.g., entropy, coefficient of variation). Results are reported in a CSV file. The evaluation framework can be further customized with post-processing (e.g., fill holes, remove small objects, thresholding using uncertainty). It is also possible to compute metrics for specific object sizes (e.g., small vs. large lesions). \texttt{ivadomed} has a module to find the optimal threshold value on the output soft prediction, via a grid-search finding applied to evaluation metrics or ROC curve.

Multiple visualization tools are included in \texttt{ivadomed} to support the design and optimization of tailored training models: GIF animations across training epochs, visual quality control of data augmentation, training curve plots, integration of the TensorBoard module, and output images with true/false positive labels. 

\section{Usage}

\paragraph{Use case 1 - Spinal Cord Toolbox:}
\href{http://spinalcordtoolbox.com/}{Spinal cord toolbox} (SCT) is an open-source analysis software package for processing MRI data of the spinal cord \citep{sct}. \texttt{ivadomed} is SCT's backbone for the automated segmentation of the spinal cord, gray matter, tumors, and multiple sclerosis lesions, as well as for the labeling of intervertebral discs.

\paragraph{Use case 2 - Creation of anatomical template:}
\texttt{ivadomed} was used in the generation of several high-resolution anatomical MRI templates \citep{calabrese_postmortem_2018, gros_ex_2020}. To make anatomical templates, it is sometimes necessary to segment anatomical regions, such as the spinal cord white matter. When dealing with high resolution data, there may be several thousand 2D slices to segment, stressing the need for a fully-automated and robust solution. In these studies, only a handful of slices were manually-segmented and used to train a specific model. The model then predicted reliably and with high accuracy (Dice score \textgreater 90\%) the delineation of anatomical structures for the thousands of remaining unlabeled slices.

\begin{figure}[H]
\vspace{-5pt}
\includegraphics[width=1\linewidth]{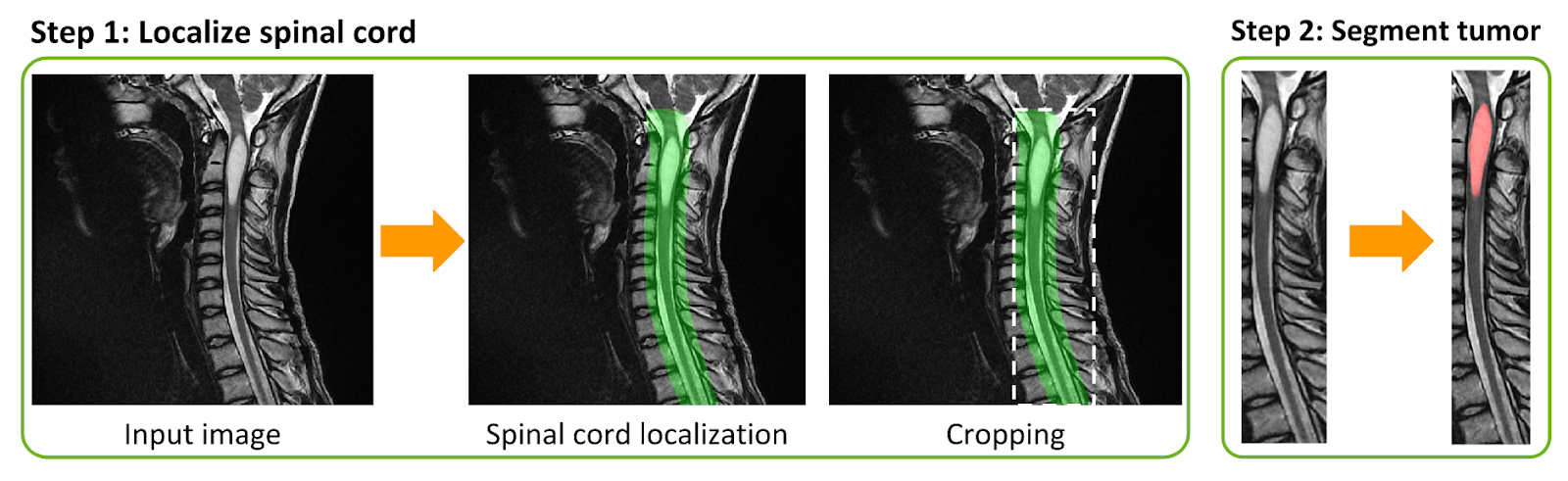}
\vspace{-20pt}
\caption{Fully automatic spinal cord tumor segmentation framework. Step 1: The spinal cord is localized using a 3D U-Net and the image is cropped around the generated mask. Step 2: The spinal cord tumors are segmented.}
\label{fig:tumor}
\vspace{-25pt}
\end{figure}

\paragraph{Use case 3 - Tumor segmentation:}
\texttt{ivadomed} also proves to be useful in the context of clinical radiology routine \citep{lemay_fully_2020}, where clinicians need to segment tumors, edema, and cavity to establish prognosis and monitor the outcome. The framework is composed of a cascaded architecture that detects the spinal cord, crops the image around the region of interest, and segments the tumor (Figure \ref{fig:tumor}). The resulting model can be applied to new data using only CPUs, which is more convenient in the clinical setting. The advanced features and architectures available in \texttt{ivadomed}, such as FiLM, were pivotal in obtaining encouraging results despite the difficulty of the task and the relatively low number of images.

\setlength{\abovecaptionskip}{10pt}




\acks{The authors thank Alexandru Jora, Nick Guenther, Christian Perone, Valentine Louis-Lucas, Benoît Sauty-De-Chalon, Alexandru Foias, Marie-Hélène Bourget and Leander Van Eekelen for their useful contributions, and Guillaume Dumas for proof-reading the manuscript. Funded by the Institut de valorisation des données (IVADO), the Canada Research Chair in Quantitative Magnetic Resonance Imaging [950-230815], the Canadian Institute of Health Research [CIHR FDN-143263], the Canada Foundation for Innovation [32454, 34824], the Fonds de Recherche du Québec - Santé [28826], the Natural Sciences and Engineering Research Council of Canada [RGPIN-2019-07244]. FRQNT Strategic Clusters Program (2020‐RS4‐265502 ‐ Centre UNIQUE  ‐Union Neurosciences \& Artificial Intelligence –Quebec, Canada First Research Excellence Fund through the TransMedTech Institute. C.G has a fellowship from IVADOMED [EX-2018-4], A.L. has a fellowship from NSERC and FRQNT, O.V. has a fellowship from NSERC, FRQNT and UNIQUE.}


\bibliography{ivadomed}

\end{document}